# Preparation and magnetic properties of $(Ln_{0.2}La_{0.2}Nd_{0.2}Sm_{0.2}Eu_{0.2})MnO_3$ (Ln = Dy, Ho, Er) high-entropy perovskite ceramics containing heavy rare earth elements[*]

QIN Jiedong[1], FENG Xingmin[2], WEN Zhiqin[1, *], TANG Li[1], LONG Defeng[1], ZHAO Yuhong[3]

1. Guangxi Key Laboratory of Optoelectronic Materials and Devices, School of Materials Science and Engineering, Guilin University of Technology, Guilin 541004, China
2. Department of Special Technology, Army Special Operations Academy, Guilin 541002, China
3. School of Materials Science and Engineering, North University of China, Taiyuan 030051, China

**Abstract**

Equimolar ratio high-entropy perovskite ceramics (HEPCs) have attracted much attention due to their excellent magnetization intensity. To further enhance their magnetization intensities, $(Ln_{0.2}La_{0.2}Nd_{0.2}Sm_{0.2}Eu_{0.2})MnO_3$ (Ln = Dy, Ho and Er, labeled as Ln-LNSEMO) HEPCs are designed based on the configuration entropy $S_{config}$, tolerance factor $t$, and mismatch degree $\sigma^2$. Single-phase HEPCs are synthesized by the solid-phase method in this work, in which the effects of the heavy rare-earth elements Dy, Ho and Er on the structure and magnetic properties of Ln-LNSEMO are systematically studied. The results show that all Ln-LNSEMO HEPCs exhibit high crystallinity and maintain excellent structural stability after sintering at 1250 °C for 16 h. Ln-LNSEMO HEPCs exhibit significant lattice distortion effects, with smooth surface morphology, clearly distinguishable grain boundaries, and irregular polygonal shapes. The three high-entropy ceramic samples exhibit hysteresis behavior at $T$ = 5 K, with the Curie temperature $T_C$ decreasing as the radius of the introduced rare-earth ions decreases, while the saturation magnetization and coercivity increase accordingly. When the average ionic radius of A-site decreases, the interaction between their valence electrons and local electrons in the crystal increases, thereby enhancing the conversion of electrons to oriented magnetic moments under an external magnetic field. Thus, Er-LNSEMO HEPC shows a higher saturation magnetization strength (42.8 emu/g) and coercivity (2.09 kOe) than the other samples, which is attributed to the strong magnetic crystal anisotropy, larger lattice distortion $\sigma^2$ (6.52×10$^{-3}$), smaller average grain size (440.49 ± 22.02 nm), unit cell volume (229.432 Å$^3$) and A-site average ion radius (1.24 Å) of its magnet. The Er-LNSEMO HEPC has potential applications in magnetic recording materials.

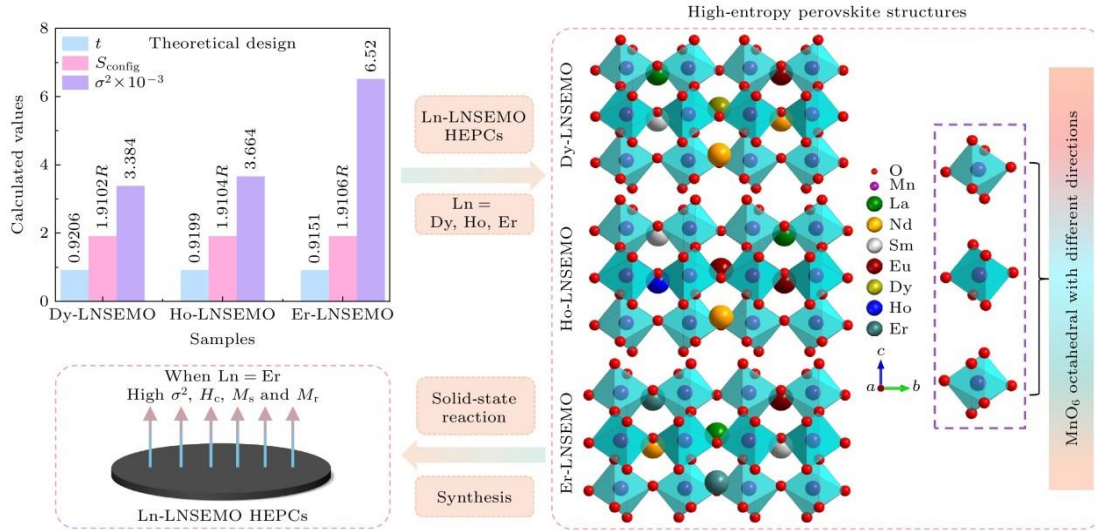

# 1. Introduction

High-entropy materials (HEMs) have become a research hotspot in the field of materials science because of their unique physical and chemical properties[1]. Initially, the study of HEMs was focused on high-entropy alloys. In 2015, Rost et al.[2] successfully synthesized (Co$_{0.2}$Cu$_{0.2}$Mg$_{0.2}$Ni$_{0.2}$Zn$_{0.2}$)O entropy-stabilized oxide. Since then, research on high-entropy ceramics (HECs) has sprung up like mushrooms, including carbides[3], nitrides[4], borides[5] and oxides[6]. Among them, the study of high-entropy oxides (HEOs) with different bonding properties and crystal structures has also developed rapidly, especially high-entropy perovskite oxides.

High-entropy perovskite oxides, owing to their diverse elements and compositions, exhibit outstanding catalytic, dielectric and magnetic properties. They hold broad application prospects in fields such as photocatalysis[7], dielectric materials[8], magnetic recording materials and sensors[9], thereby attracting significant attention from researchers. In high-entropy perovskite ceramics (HEPCs), both A-site and B-site may accommodate multi-element combinations, which endows high-entropy components with rich diversity[10]. By adjusting the average ionic radius of A-site, large lattice distortion, more defects and highly disordered lattice arrangement[11] can be generated, which makes HEPCs have excellent temperature stability and favorable magnetic properties[12]. Additionally, most rare earth ions exhibit active chemical activity due to the presence of unfilled 4$f$ shells[13]. Elements containing 4$f$ electrons show unique properties in chemical reactions by affecting the effective nuclear charge number, electronic transition, bonding ability and oxidation state stability of atoms. Rare earth elements can interact with 3$d$ transition metals, such as magnetic coupling and exchange interactions, to form a variety of compounds with unique magnetic and electronic properties[14]. Currently, the common method to adjust the magnetic properties of rare-earth transition metal compounds is to place multiple rare-earth elements with equal atomic ratio into the A-site of HEPCs through a high-entropy strategy. For example, Zhao et al.[15] designed and synthesized nanofibers $(Sm_{0.2}Eu_{0.2}Gd_{0.2}Ho_{0.2}Yb_{0.2})CrO_3$ HECs with orthorhombic perovskite phase structure using electrospinning technology, exhibiting a typical negative temperature coefficient. Krawczyk et al.[16] synthesized single-phase $(Gd_{0.2}Nd_{0.2}La_{0.2}Sm_{0.2}Y_{0.2})CoO_3$ HECs with ferromagnetism by a coprecipitation-hydrothermal method, observing a transition in their magnetic properties from a low-spin to a high-spin state. Witte et al.[17] prepared three composite systems of mixed rare earth A-site cations and single transition metal B-site cations, specifically $(Gd_{0.2}La_{0.2}Nd_{0.2}Sm_{0.2}Y_{0.2})Co(Cr, Fe)O_3$. The results show that these samples exhibit significant magnetic interaction at low temperatures. In addition, our group has designed and synthesized several single-phase HEPCs with orthorhombic perovskite structure via solid-state reaction, such as $(Gd_{0.2}La_{0.2}Nd_{0.2}Sm_{0.2}Y_{0.2})MnO_3$, $(Eu_{0.2}Gd_{0.2}La_{0.2}Nd_{0.2}Sm_{0.2})MnO_3$, $(Ho_{0.2}Gd_{0.2}La_{0.2}Nd_{0.2}Sm_{0.2})MnO_3$ and $(Yb_{0.2}Gd_{0.2}La_{0.2}Nd_{0.2}Sm_{0.2})MnO_3$[18,19]. The above studies show that high-entropy perovskite oxides composed of rare earth and transition metals exhibit a distinct single-phase structure and demonstrate outstanding properties.

Magnetic recording materials play a vital role in modern information storage technology. Ideal magnetic recording materials need to have high saturation magnetization and appropriate coercivity (0.5-3 kOe)[20] to achieve efficient, reliable, and energy-saving data storage. In recent years, HEPCs composed of rare earth elements and transition metals have provided new ideas for the development of magnetic recording materials. However, the performance regulation of HEPCs still faces challenges, especially in the optimization of magnetic properties, which has become one of the key issues to be solved urgently in the scientific community. Moreover, HEPCs may encounter issues of structural instability and performance degradation during high temperatures or long-term use. Consequently, it is necessary to conduct further systematic investigations into the crystal structure and magnetic properties of HEPCs in order to gain a more comprehensive understanding of the relationship between their microstructure and performance, as well as their magnetic mechanisms, thereby broadening their applications in magnetic recording materials.

To achieve HEPCs with superior magnetic properties, this study focuses on light rare-earth elements such as La, Nd, Sm and Eu based on the research group's prior work. By introducing heavy rare-earth ions (Dy, Ho, Er)[21] possessing high magnetic crystalline anisotropy and large magnetic moments to occupy the A site, the aim is to optimize the magnetic performance of the HEPCs. Thus, in this study, $(Ln_{0.2}La_{0.2}Nd_{0.2}Sm_{0.2}Eu_{0.2})MnO_3$ (Ln-LNSEMO) (Ln = Dy, Ho and Er) HEPCs were designed based on the theoretical calculation of configurational entropy $S_{config}$, mismatch $\sigma^2$ and tolerance factor $t$. HEPCs containing heavy rare-earth elements were prepared via the solid-state method. The effects of grain size and large magnetic moment heavy rare earth elements (Ln = Dy, Ho, Er) on the structure and magnetic properties of Ln-LNSEMO HEPCs were systematically investigated. The intrinsic mechanism linking structure and magnetism was elucidated, providing novel insights for the development of high-performance and high-entropy oxides.

## 2. Experimental method

2.1 Preparation method

Ln-LNSEMO (Ln = Dy, Ho, Er) HEPCs were synthesized by the solid-state reaction method. Commercially available $La_2O_3$, $Nd_2O_3$, $Sm_2O_3$, $Eu_2O_3$, $Dy_2O_3$, $Ho_2O_3$, $Er_2O_3$ (99.99%) and $MnO_2$ (85.00%) powders, which are provided by Sinopharm Chemical Reagent Co. Ltd, China, were used as raw materials to form a series of single-phase perovskite manganese oxides. The powders were weighed according to stoichiometry and poured into an agate jar containing 20 mL of absolute ethanol as the ball milling medium. Then, the mixture was milled in a planetary ball mill (YXQM-4L, Changsha Mi Qi Together Equipment Co., Ltd) for 13 h at a speed of 250 r/min to ensure it was thoroughly mixed. After grinding, the suspension was placed in a blast drying oven and dried at a constant temperature of 80 °C until the absolute ethanol was completely volatilized. Grind the dried powder in an agate mortar for 30 min to ensure uniformity. Subsequently, it was calcined in a high-temperature muffle furnace (KSL-1700 X-A1, Hefei Kejing Material Technology Co., Ltd.) at 1200 °C for 16 h to obtain the precursor. The powder was compressed into a block-shaped sample under a unidirectional pressure of 20 MPa and held at that pressure for 10 min. Then, the bulk ceramic sample was put into an alumina crucible and heated to the target temperature of 1250 °C at a rate of 5 °C/min for 16 h (heating for 4 h and holding for 12 h), and then cooled in the furnace to obtain single-phase Ln-LNSEMO HEPCs.

2.2 Characterization

The samples were characterized by X-ray diffraction (XRD, X′ Pert PRO) with Cu Kα radiation in the scan range of 20°–80° at a scan rate of 2 °/min. The Rietveld refinement technique in FullProf software was used to refine the XRD patterns of all samples. The elemental composition and micromorphology of Ln-LNSEMO (Ln = Dy, Ho, Er) samples were analyzed and observed by scanning electron microscopy (SEM, HITACHI S-4800) equipped with an energy dispersive X-ray spectroscopy (EDS) detector. The Al Kα target was characterized by excited X-ray photoelectron spectroscopy (XPS, ESCALAB-250 Xi), performed at a $10^{-8}$ Pa test pressure with energy resolution ⩽ 0.45 eV and spatial resolution ⩽ 20 μm. The magnetic properties of the samples were measured by a commercial superconducting quantum interference device (SQUID, MPMS3, Quantum Design) and a vibrating sample magnetometer (VSM). The instrument has an AC susceptibility frequency range of 0.1 Hz–1000 Hz, a maximum measured magnetic moment of 10 emu, and an applied magnetic field strength range of $H = \pm 70$ kOe. In this work, an external magnetic field strength in the range of $H = \pm 30$ kOe was applied to all samples to measure the magnetization versus temperature curve (M-T) and the magnetization versus applied field curve (M-H).

## 3. Results and discussion

3.1 Phase composition and microstructure

The configurational entropy $S_{config}$, the mismatch $\sigma^2$ and the tolerance factor $t$ are important indicators for designing Ln-LNSEMO (Ln = Dy, Ho, Er) HEPCs. The mixture of sublattice components formed by group IV transition metals and rare earth elements in multicomponent oxides leads to their high stability[22]. The concept of entropy stability is based on the fact that the increase of the configurational entropy of the system makes the Gibbs free energy of the system decrease, thus stabilizing the single-phase crystal structure. The configurational entropy of a multicomponent mixture is calculated as follows:

$$S_{config} = -R \left[ \left( \sum_{i=1}^{M} x_i \ln x_i \right)_{A\text{-site}} + \left( \sum_{j=1}^{N} x_j \ln x_j \right)_{B\text{-site}} + 3 \left( \sum_{k=1}^{P} x_k \ln x_k \right)_{O\text{-site}} \right] \quad (1)$$

Where $R$ represents the ideal gas constant; $M$, $N$ and $P$ are the number of ions at A, B and O sites, respectively; $x_i$, $x_j$ and $x_k$ are the mole fractions of A, B and O site ions, respectively. The degree of mismatch $\sigma^2$ is one of the important parameters of perovskite manganese oxide. It is defined as follows:

$$\sigma^2 = \left( r_A^2 \right) - \left( r_A \right)^2 = \sum x_i r_i^2 - \left( \sum x_i r_i \right)^2 \quad (2)$$

Where $x_i$ and $r_i$ are the proportion of rare earth ions in the A-site and the size of the ionic radius, respectively, where the ionic radius of each element is listed in Tab. 1. The size of the radius of rare earth ions is Dy> Ho>Er, so the average ionic radius of A-site in the sample decreases, as shown in Fig. 1(a). In addition, the calculated results of $S_{config}$ and $\sigma^2$ of Ln-LNSEMO (Ln = Dy, Ho, Er) HEPCs are displayed in Fig. 1(b). According to Fig. 1(b), the $S_{config}$ of the synthesized ceramic samples is all greater than $1.5R$, which belongs to the high-entropy category[23], indicating that the designed HEPCs have high configurational entropy.

**Table 1.** Oxidation state, co-ordination number (CN) and corresponding ionic radius ($r$)[23].

| Element | Oxidation | CN  | $r$/Å |
|---------|-----------|-----|-------|
| La      | 3+        | XII | 1.36  |
| Nd      | 3+        | XII | 1.27  |
| Sm      | 3+        | XII | 1.24  |
| Eu      | 3+        | XII | 1.22  |
| Dy      | 3+        | XII | 1.19  |
| Ho      | 3+        | XII | 1.18  |
| Er      | 3+        | XII | 1.11  |
| Mn      | 3+        | VI  | 0.64  |
| O       | 2−        | VI  | 1.40  |

The Goldschmidt tolerance factor $t$ helps to determine whether a material can form a stable perovskite structure[24]. The equation for calculating the tolerance factor $t$ in terms of the ionic radius is defined as follows[25]:

$$t = \frac{r_A + r_O}{\sqrt{2}(r_B + r_O)} \qquad (3)$$

Where $r_A$, $r_B$, and $r_O$ are the average ionic radii of A, B and O sites in perovskite oxides, respectively. In perovskite manganates, the stable perovskite structure can be maintained when the tolerance coefficient $t$ is between 0.78 and 1.05[26]. The $t$ value of Ln-LNSEMO HEPCs is calculated to decrease with decreasing ionic radius (see Fig. 1(b)). Moreover, the tolerance factor of the designed system is close to 1, which indicates that the system has good stability.

Fig. 1(c)-(f) are XRD patterns and Rietveld refinement maps of Ln-LNSEMO HEPCs. Observation of Fig. 1(c) reveals that the XRD patterns of all HEPCs exhibit sharp and distinct diffraction peaks, and there is no second phase. This confirms that HEPCs possess a single perovskite structure with excellent crystallinity. The diffraction peaks of all samples match well with the diffraction data of single rare-earth manganese oxides $SmMnO_3$, $NdMnO_3$ and $LaMnO_3$. Furthermore, the main diffraction peaks of the samples corresponded to the main diffraction peaks

of EuMnO$_3$, DyMnO$_3$, HoMnO$_3$ and ErMnO$_3$. The results show that the rare-earth elements La, Nd, Sm, Eu and Dy (or Ho, or Er) are located at the A-site of Ln-LNSEMO HEPCs in an equimolar ratio. The XRD data of the prepared Ln-LNSEMO HEPCs were refined by Rietveld based on PDF standard card: 25-0747 (SmMnO$_3$), as shown in Fig. 1(d)-(f). When the important factors $R_p$, $R_{wp}$ and $R_{exp}$ are all less than 15%[27], it can be observed that the original XRD data and the calculated fitting data show good agreement. According to the fitting results, the $R_p$, $R_{wp}$ and $R_{exp}$ of all samples are less than 5.5%. In addition, the $\chi^2$ values of all samples are less than 1, indicating that the refined fitting results for the synthesized Ln-LNSEMO HEPCs exhibited highly reliable[28]. The refinement results further demonstrated that all samples belonged to the orthorhombic structure with space group *Pbnm*.

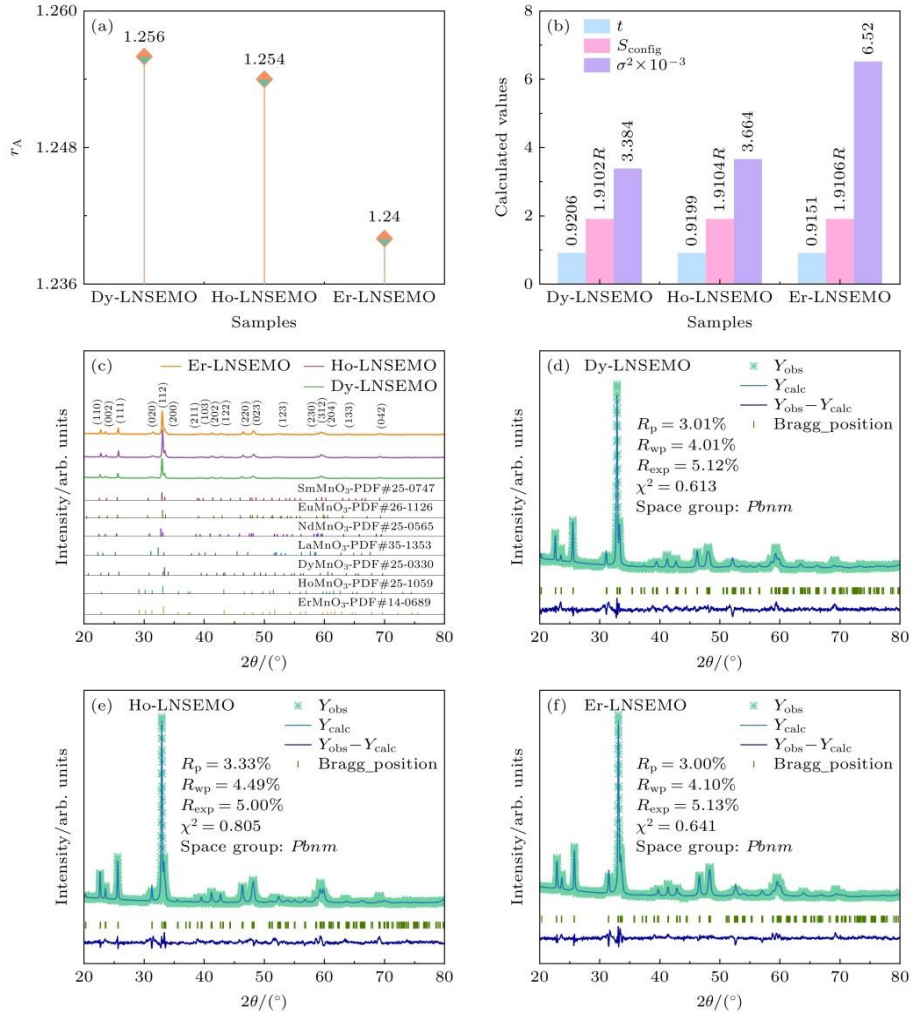

**Figure 1.** Characteristics of Ln-LNSEMO ceramics sintered at 1250 °C: (a) Average ionic radius of A-site; (b) theoretical calculation values of tolerance factor $t$, configuration entropy $S_{config}$ and mismatch degree $\sigma^2$; (c) X-ray diffraction patterns; (d)–(f) Rietveld refinement.

Fig. 2(a), (b) shows the lattice parameters and unit cell volume after Rietveld refinement of XRD data. The lattice parameters of the three components are similar, but the unit cell volume decreases with the decrease of the radius of the heavy rare earth ion, which leads to the increase of the distortion and deformation in the lattice. When the crystal is formed, the larger Dy$^{3+}$ ions will occupy a larger space in the lattice, which makes the whole unit cell expand outward and further increases the volume of the unit cell. With the decrease of the radius of heavy rare earth ions (Dy, Ho, Er), the difference of electronegativity between heavy rare earth ions and light rare earth ions (La, Nd, Sm, Eu) increases. This indicates that the distortion effect between light and heavy rare earth ions and the coupling effect between rare earth ions and manganese ions are enhanced. At the same time, the decrease of the tolerance factor $t$ of the designed high-entropy component indicates that the gap between the average ionic radii of A-site and B-site increases, which leads to the increase of the atomic layer mismatch between A-O layer and B-O layer, thus causing lattice distortion. However, the mismatch $\sigma^2$ of the samples increases with the introduction of small ionic radius rare earth elements (Fig. 1(b)), which to some extent reflects that the increase of ionic radius difference between A-site elements in the perovskite structure enhances the lattice distortion of the materials, thus affecting the Curie temperature and magnetic properties of HEPCs. In addition, the Fig. 2(c) cell model shows a high degree of atomic disorder, lattice distortion, and manganese-oxide octahedral distortion patterns, which are attributed to differences in atomic size, mass, and electronegativity. This confirms the lattice distortion effect of Ln-LNSEMO HEPCs.

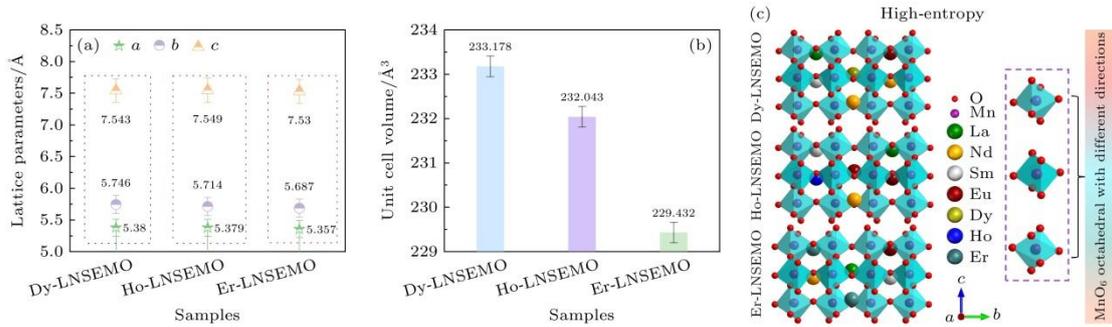

**Figure 2.** Lattice parameters (a), cell volume (b) of Ln-LNSEMO ceramics after Rietveld refinement and crystal structure of samples (c).

Tab. 2 lists the bond length $d$ and bond angle $\theta$ of the refined Ln-LNSEMO HEPCs. The introduction of rare-earth elements with different radii affects the octahedral structure of $MnO_6$ in perovskite manganese oxide, destroys the periodic arrangement in the lattice and results in lattice distortion, which is manifested by the changes of Mn-O bond length and Mn-O-Mn bond angle. Small changes in these two parameters can significantly affect the magnetic properties of Ln-LNSEMO, reflecting the sensitivity and complexity of the manganese oxide octahedral structure. Among them, the Er-LNSEMO sample has the longest Mn-O bond length and the largest Mn-O-Mn bond angle, which may alter the atomic equilibrium force in the lattice, leading to a decrease in the stability of the lattice structure. However, these significant changes can be attributed to the tilt of Jahn-Teller (JT) distortion (all compounds satisfy the condition of JT distortion [$b > a > c/\sqrt{2}$])[29]. Therefore, these distortions may further affect the electronic structure and redox reactions of HEPCs, thereby affecting the magnetic behavior of the sample.

Table 2. Bond length $d$ and bond angle $\theta$ of three groups of Rietveld refined samples.

| Samples | $d_{Mn-O}$ /Å | $\theta_{Mn-O-Mn}$/(°) |
|---|---|---|
| Dy-LNSEMO | 1.9157(3) | 148.167(6) |
| Ho-LNSEMO | 1.9357(2) | 141.748(6) |
| Er-LNSEMO | 1.9500(3) | 152.118(4) |

To observe the microstructure of the sample and verify the accuracy of the composition design, the surface morphology, particle size distribution, EDS elemental map and chemical composition of Ln-LNSEMO HEPCs were analyzed in detail by SEM and EDS, as shown in Fig. 3. During the sintering reaction, grains gradually grow from their nuclei and tend to approach each other, ultimately forming grain boundaries. The orientation of these grain boundaries is disordered, resulting in inconsistencies in grain shape and size, as well as diversity in grain boundary angles. Therefore, the SEM images of all samples show smooth surfaces, clear grain boundaries and irregular polygonal shapes, as shown in Fig. 3(a)-(c). In Ln-LNSEMO HEPCs, the number of grains in the size range of 400-500 nm is higher than that in other size ranges, indicating a higher trend of grain formation and growth within this size range. As shown in the inset of Fig. 3(a)-(c), the average grain size (AGS) of Dy-LNSEMO, Ho-LNSEMO and Er-LNSEMO is (499.98 ± 24.99) nm, (474.83 ± 23.74) nm and (440.49 ± 22.02) nm, respectively. Therefore, it can be concluded that AGS shows a decreasing trend. The EDS elemental distribution map shows that all elements are uniformly distributed without any elemental segregation, which further proves that rare-earth elements are located at the A-site of Ln-LNSEMO HEPCs. Fig. 3(a1)-(c1) is the atomic percentage of the chemical composition of the relevant element in all samples. The atomic percentages of La, Nd, Sm, Eu, Dy, Ho and Er are similar. At the same time, the atomic percentage ratio of O to Mn is approximately 3:1, which is close to the designed stoichiometric ratio, which confirms the accuracy of the composition design.

The valence states and related chemical information of Mn and O in Ln-LNSEMO HEPCs were analyzed by XPS, as shown in Fig. 4. To achieve high-quality peak fitting and ensure accurate representation and reliable results of the spectra, all core-level spectra were fitted using Gaussian Lorentzian peaks combined with Shirley background[30]. Two different peaks were detected in the O 1s spectrum of Ln-LNSEMO HEPCs, corresponding to lattice oxygen ($O_L$) and defect oxygen ($O_V$), respectively[31]. The percentage of $O_V$ ($O_L$) in Dy-LNSEMO, Ho-LNSEMO and Er-LNSEMO HEPCs is 57% ± 2.85% (43% ± 2.15%), 66% ± 3.3% (34% ± 1.7%) and 53% ± 2.65% (47% ± 2.35%), respectively. Ho-LNSEMO has the highest content of $O_V$. In addition, the XPS fine spectra of O 1s orbital of all samples at room temperature are shown in Fig. 4(a). From the figure, it can be seen that the positions of $O_V$ in the samples of Dy-LNSEMO, Ho-LNSEMO and Er-LNSEMO HEPCs correspond to the peaks binding energy positions of 530.36 eV, 529.71 eV and 530.07 eV, respectively. While the peaks at 528.76 eV, 528.29 eV and 528.51 eV correspond to the $O_L$ of Ln-LNSEMO HEPCs. In addition, from Fig. 4(b), it can be seen that the Mn 2p peak of all samples is divided into two regions, corresponding to the spectra of Mn $2p_{1/2}$ and Mn $2p_{3/2}$ regions, respectively.

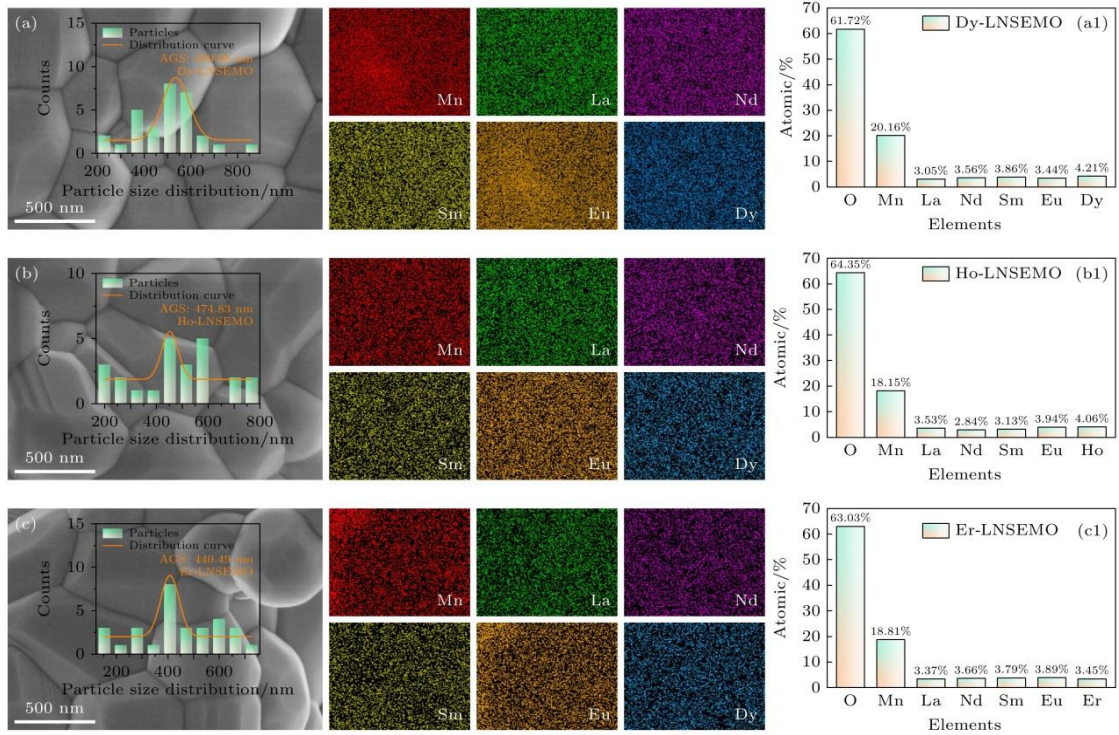

**Figure 3.** SEM micrographs, particle size distribution, EDS mapping and chemical composition (at. %) of Ln-LNSEMO HEPCs sintered at 1250 °C: (a), (a1) Dy-LNSEMO; (b), (b1) Ho-LNSEMO; (c), (c1) Er-LNSEMO.

The peaks of 652.75 eV & 640.57 eV, 652.37 eV & 640.49 eV, and the peaks at 652.55 eV & 640.67 eV correspond to $Mn^{3+}$ in Mn $2p_{1/2}$ and Mn $2p_{3/2}$ orbitals of Ln-LNSEMO, respectively. The peak at 641.9 eV corresponds to $Mn^{4+}$ in the Mn $2p_{3/2}$ spectrum of Dy-LNSEMO HEPCs, while the peak at 642.56 eV (656.04 eV) corresponds to $Mn^{4+}$ in the Mn $2p_{3/2}$ (Mn $2p_{1/2}$) spectrum of Er-LNSEMO HEPCs. However, the presence of $Mn^{4+}$ was not detected in the Ho-LNSEMO sample. This is because during the reduction reaction of this sample, almost all $Mn^{4+}$ is reduced to $Mn^{3+}$, resulting in more defective oxygen. In addition, the manganese ions in Dy-LNSEMO and Er-LNSEMO samples exhibit a mixed valence state, resulting in $Mn^{3+}$-$O^{2-}$-$Mn^{4+}$ double exchange interaction. The proportion of $Mn^{4+}$ ($Mn^{3+}$) in Dy-LNSEMO and Er-LNSEMO is 28% ± 1.4% (72% ± 3.6%) and 26% ± 1.3% (74% ± 3.7%), respectively, with differences of 44% ± 2.2% and 48% ± 2.4%, respectively. The larger the difference between $Mn^{4+}$ and $Mn^{3+}$, the more it can promote electron transfer between transition metals of different valence states. It can be seen that Er-LNSEMO has the strongest double exchange interaction in all samples. According to Fig. 4(c) and (d), the metallic states of rare-earth elements Dy, Ho and Er in the synthesized ceramic samples are trivalent[32,33], which is consistent with the designed valence states of rare-earth elements.

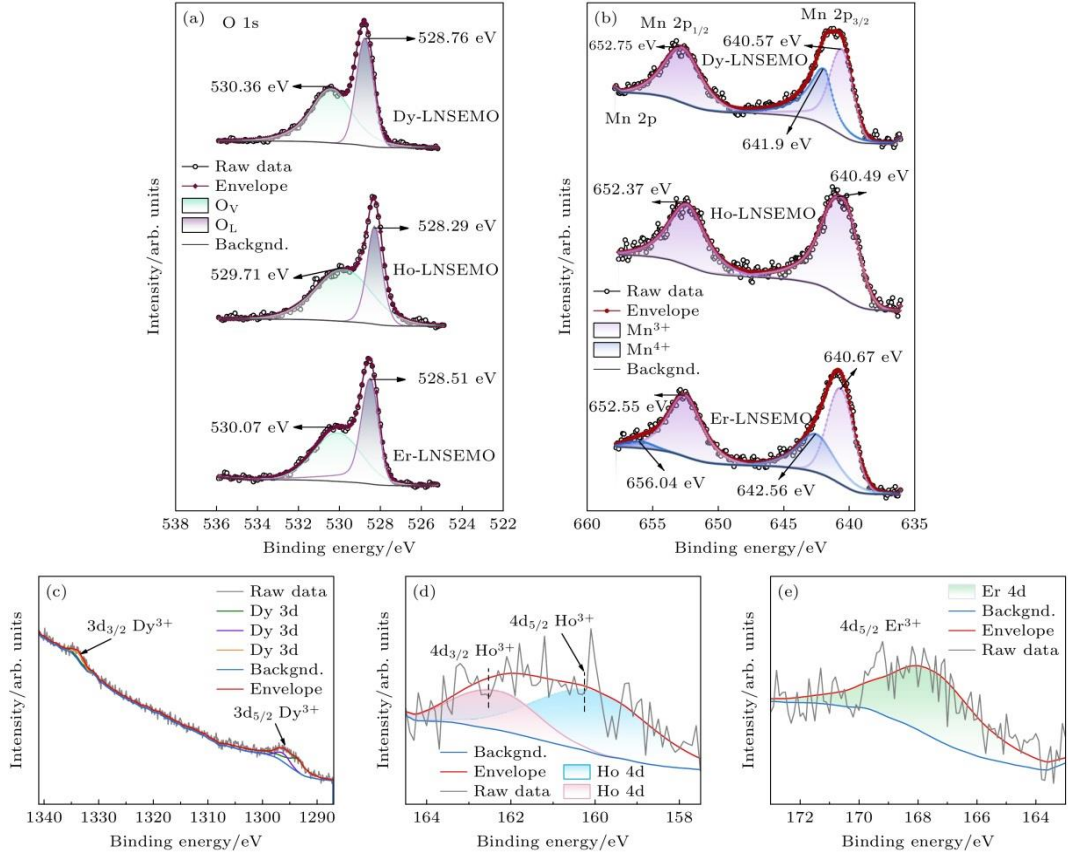

**Figure 4.** High-resolution XPS spectra of Ln-LNSEMO HEPCs: (a) O 1s; (b) Mn 2p; (c) Dy 3d; (d) Ho 4d; (e) Er 4d.

3.2 Magnetic property

To evaluate the Curie temperature of Ln-LNSEMO HEPCs, the magnetization versus temperature curves (*M-T*) of all samples were tested under a magnetic field of 0.1 kOe and a temperature range of 30−80 K, as shown in Fig. 5(a). With the increase of test temperature, the magnetization intensity of Ln-LNSEMO gradually decreases, and the curve tends to flatten with a slope close to zero, indicating that the sample has undergone magnetic transformation within this temperature range. In addition, the Curie temperature ($T_C$) of the sample is determined by measuring the derivative of the magnetization with temperature (d*M*/d*T*) (i.e., differentiating the *M-T* curve). As shown in Fig. 5(b), the abscissa corresponding to the unique extreme point in the d*M*/d*T*-*T* curve represents the $T_C$ of the sample, at which a transition from ferromagnetism (FM) to paramagnetism (PM) occurs[34]. As shown in the figure, the $T_C$ of the sample decreases with the decrease of the radius of the introduced heavy rare earth ions. This phenomenon is attributed to the lattice distortion of Ln-LNSEMO, which is mainly manifested by the tolerance factor *t* and the mismatch $\sigma^2$. Specifically, $\sigma^2$ increases significantly with the decrease of the radius of the doped heavy rare-earth ion, while the tolerance factor *t* shows the opposite trend (see Fig. 1(b)), indicating an increase in the degree of lattice distortion. Additionally, the *t* and $\sigma^2$ factors affect the

MnO$_6$ octahedron, which contains the single electron bandwidth $W$[35]. The equation for $W$ is as follows:

$$W = \frac{\cos\left(\frac{\pi - \theta_{\text{Mn-O-Mn}}}{2}\right)}{(d_{\text{Mn-O}})^{3.5}} \quad (4)$$

Where $\theta_{\text{Mn-O-M}}$ and $d_{\text{Mn-O}}$ are the Mn-O-M bond angle and Mn-O bond length, respectively. After calculation, the bandwidth $W$ of Ln-LNSEMO HEPCs decreases from 0.09882 to 0.09373. The decrease of $W$ indicates a reduction in the overlap between Mn 3d and O 2p orbitals, which enhances the electron-phonon coupling effect[36] and finally leads to a decrease in the magnetic transition temperature $T_C$.

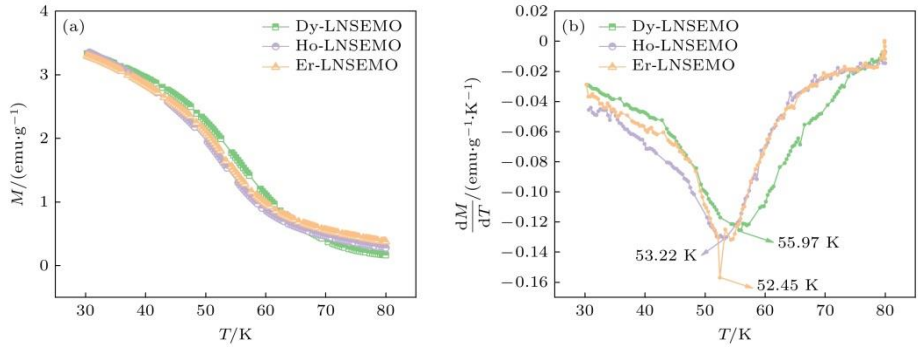

**Figure 5.** $M$-$T$ curves (a) and d$M$/d$T$-$T$ curves (b) of Ln-LNSEMO HEPCs.

To confirm the degree of magnetization of the sample under the action of magnetic field, the hysteresis loop of Ln-LNSEMO HEPCs was tested at $T$ = 5 K, as shown in Fig. 6(a), with an applied external magnetic field range of -30 kOe to 30 kOe. From the illustration in Fig. 6(a), it can be seen that the saturation magnetization of Er-LNSEMO is significantly higher than that of other samples. Moreover, the slope of the magnetization curve of all samples gradually decreases when the maximum external magnetic field is applied. The weak coupling effect between the magnetic moment of rare-earth ions and the Mn sublattice, or some highly suppressed magnetic exchange interactions, can cause a unique trend in the magnetic ordering state inside the material, greatly affecting the magnetic properties of the material under different temperature and external magnetic field conditions[17]. This indicates that certain mechanisms within HEPCs may hinder the complete alignment of magnetic moments, resulting in the slope of the magnetization curve of the sample not close to zero under high magnetic field. Fig. 6(b) shows magnetization curves in the low magnetic field range of -3 kOe to 3 kOe, indicating that all Ln-LNSEMO HEPCs exhibit pronounced hysteresis behavior.

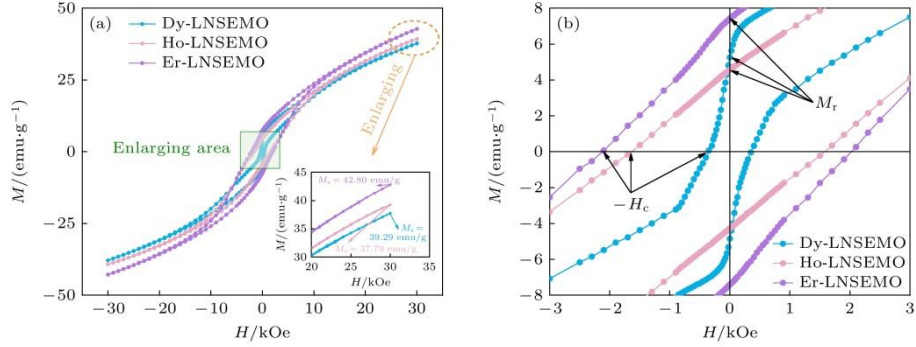

**Figure 6.** Hysteresis loops at $T$ = 5 K (a), the magnified magnetization curves in the low magnetic field region (b) of Ln-LNSEMO HEPCs.

Fig. 7 is the magnetic properties parameters of the sample, including saturation magnetization ($M_s$), remanent magnetization ($M_r$) and coercivity ($H_c$). These parameters can reflect the magnetization characteristics and magnetic stability of Ln-LNSEMO HEPCs. As the average ionic radius of A-site decreases, the $M_s$ and $H_c$ of Ln-LNSEMO HEPCs increase, while the $M_r$ first decreases and then increases. It is attributed to the spin-orbit coupling effect between the rare-earth ions and manganese ions within HEPCs, the interaction of 4$f$ electrons of rare-earth ions, and the enhancement of magnetic crystal anisotropy, which further enhances the magnetization characteristics of HEPCs[37]. The magnetic crystal anisotropy originates from the unrestricted orbital angular momentum of 4$f$ electron of the rare-earth ions; The electronic interaction between rare-earth ions is the interaction of 4$f$ electron spins between different rare-earth ions, which affects the spin arrangement of unpaired electrons and thus affects the magnetism of HEPCs.

Moreover, as the radius of heavy rare-earth ions decreases, the covalent properties of HEPCs also increase, leading to an increase in the coordination ability between manganese ions and rare-earth ions[21]. This enhances the magnetic anisotropy of Ln-LNSEMO HEPCs while significantly increasing their energy barrier, thereby affecting the magnetic properties of HEPCs. In addition, Er-LNSEMO HEPCs have the largest $M_s$, $M_r$, and $H_c$. The higher $H_c$ can be attributed to the strong magnetic crystal anisotropy of the magnet, resulting in greater hysteresis behavior during magnetization. Additionally, the remanence ratios ($M_r/M_s$) of Dy-LNSEMO, Ho-LNSEMO, and Er-LNSEMO HEPCs are 0.13, 0.12 and 0.17, respectively. The higher the $M_r/M_s$ ratio, the more likely the material is to maintain its magnetic field state, which in turn means that the material has better magnetic properties[38]. Accordingly, the magnetic properties of Er-LNSEMO HEPCs are the best among the three samples. Compared with other light rare-earth perovskite compounds $ABO_3$[39–41] (for example, under the same testing conditions, the residual magnetization and coercivity of $LaCoO_3$ and $La_{0.7}Ag_{0.3}CoO_3$ compounds are 0.801 emu/g (4.45 kOe) and 1.180 emu/g (7.259 kOe), respectively), it can be inferred that all Ln-LNSEMO HEPCs exhibit excellent magnetic properties. Er-LNSEMO HEPCs have high $M_s$ (42.8 emu/g), $M_r$ (7.43 emu/g), and a suitable $H_c$ (2.09 kOe), meeting the requirements of magnetic recording materials and demonstrating their potential for application in magnetic recording.

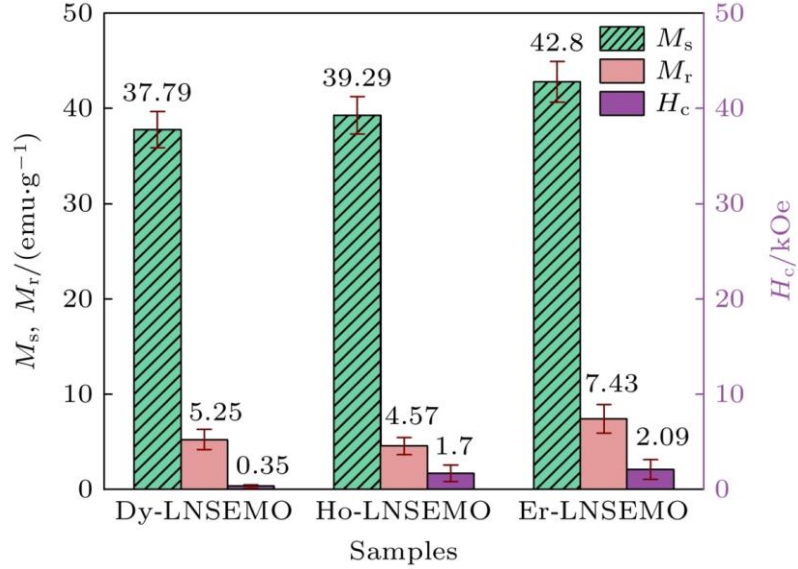

**Figure 7.** Magnetic properties parameters of Ln-LNSEMO HEPCs.

## 4. Conclusion

In this paper, Ln-LNSEMO (Ln = Dy, Ho, Er) was designed based on $S_{config}$, $t$ and $\sigma^2$, and three kinds of HEPCs were prepared by the solid-state method. The results indicate that all samples exhibit a single-phase orthorhombic perovskite structure with irregular polygonal shapes and distinct grain boundaries. The interaction of 4$f$ electron spin of rare-earth ions, grain size and $MnO_6$ octahedral distortion affect the magnetic interaction of HEPCs. Due to the large lattice distortion, the $T_C$ of all Ln-LNSEMO decreases as the radius and bandwidth $W$ of the introduced rare-earth ions decrease. When the radius of heavy rare-earth ions decreases, the interaction between its valence electrons and local electrons in the crystal increases, thereby enhancing the conversion of electrons to oriented magnetic moments under an external magnetic field. Therefore, all Ln-LNSEMO HEPCs exhibit excellent magnetic properties compared to non-heavy rare-earth oxides. Among them, Er-LNSEMO has a smaller average grain size and unit cell volume, as well as larger $\sigma^2$, $MnO_6$ octahedral distortion and stronger double exchange interaction, which makes it show high $M_s$ (42.8 emu/g), $M_r$ (7.43 emu/g) and $H_c$ (2.09 kOe).